\documentclass[10pt,conference]{IEEEtran}
\usepackage{cite}
\usepackage{epsfig}
\usepackage{times}
\usepackage{graphicx}
\usepackage[figuresright]{rotating}
\usepackage{subfigure}
\usepackage{amssymb}
\usepackage{amsmath}
\usepackage{setspace}
\usepackage{rotating}
\usepackage{multirow}
\usepackage{wrapfig}
\usepackage{booktabs}
\usepackage{array}
\usepackage{comment}
\usepackage{subfigure}
\usepackage{epstopdf}
\IEEEoverridecommandlockouts

\usepackage{algorithmic}
\usepackage{algorithm}
\floatname{algorithm}{Algorithm} \textfloatsep 0.5cm

\usepackage{colortbl}

\begin{document}
\title{Low-power All-analog Circuit for Rectangular-type Analog Joint Source Channel Coding\vspace{-0.7cm}}
\author{{\bf Xueyuan Zhao, Vidyasagar Sadhu, and Dario Pompili}\\
Department of Electrical and Computer Engineering, Rutgers University--New Brunswick, NJ, USA\\
E-mails: xueyuan\_zhao@cac.rutgers.edu, vidyasagar.sadhu@rutgers.edu, pompili@cac.rutgers.edu}
\maketitle

\thispagestyle{empty}
\pagestyle{empty}

\begin{abstract}
A low-complexity all-analog circuit is proposed to perform efficiently Analog Joint Source Channel Coding (AJSCC), which can compress two or more sensor signals into one with controlled distortion while also being robust against wireless channel impairments.
The idea is to realize the rectangular-type AJSCC using Voltage Controlled Voltage Sources (VCVS). The proposal is verified by Spice simulations as well as breadboard and Printed Circuit Board~(PCB) implementations. Results indicate that the design is feasible for low-complexity systems like persistent wireless sensor networks requiring low circuit power.

\end{abstract}

\section{Introduction}\label{sec:introduction}
\begin{figure}
\begin{center}
\includegraphics[width=3.2in]{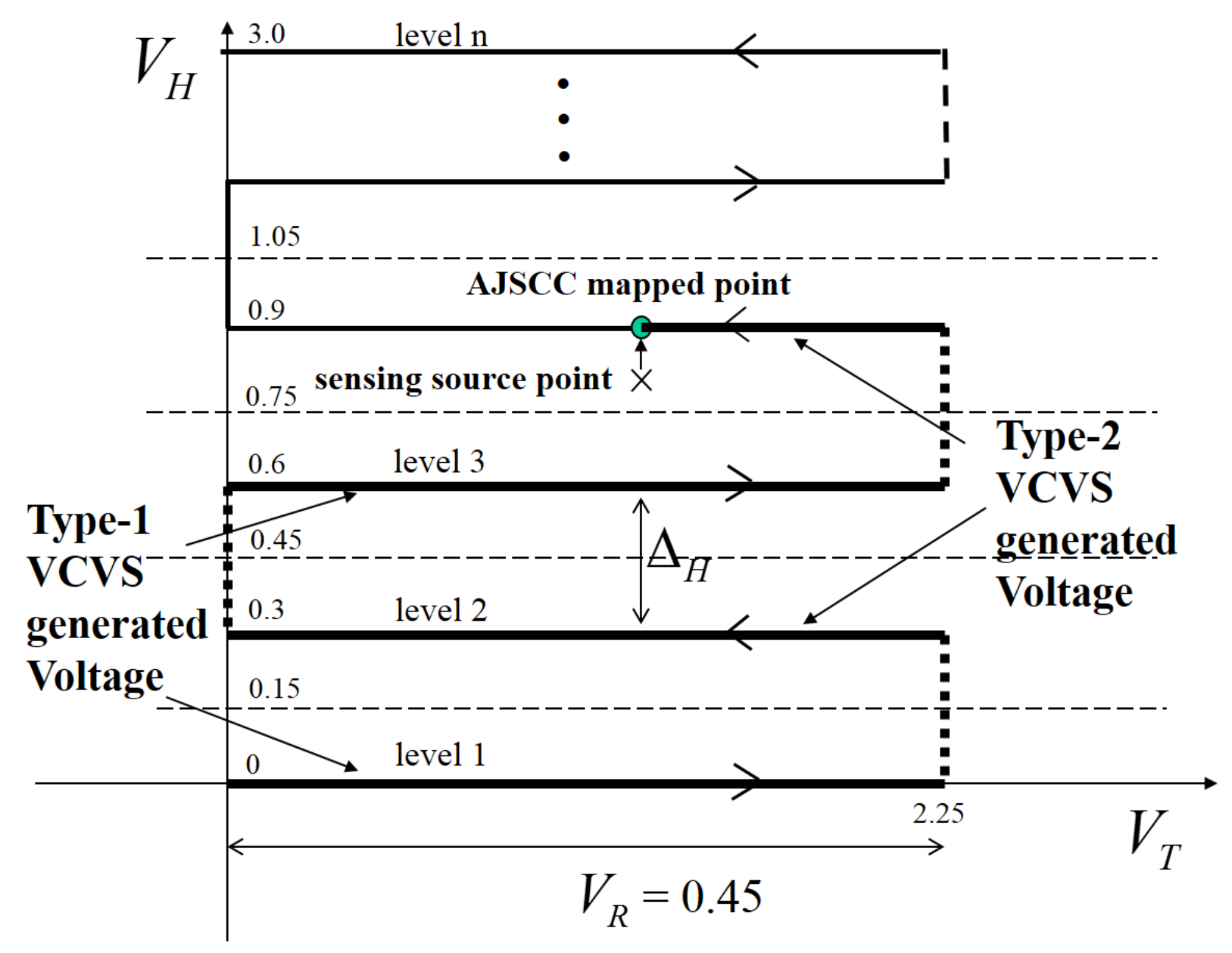}
\end{center}
\caption{\textbf{Shannon's Rectangular Mapping.} The sensed point is mapped to the point closest on the rectangular curve, and the accumulated length of the curve from the origin to the mapped point (in bold) is transmitted instead of the two values identifying the 2D sensed point. In our novel all-analog implementation, odd-level voltages are generated using Type-1 Voltage Controlled Voltage Sources~(VCVS), whereas even-level voltages are generated using Type-2 VCVS. $V_H$ and $V_T$ are the humidity and temperature variable voltages, respectively, whereas $V_R$ is a constant and $\Delta_H$ is the spacing between the levels. The numbers on the figure are the voltages used in Spice simulations.}\label{fig:rec_mapping}
\end{figure}

Analog Joint Source Channel Coding~(AJSCC)~\cite{Hekland05} can compress two or more sensor signals into one with controlled distortion while also being robust against wireless channel impairments. AJSCC adopts Shannon mapping as its encoding method~\cite{Fresnedo15}. Such mapping, in which the design of \emph{rectangular (parallel) lines} can be used for 2:1 compression (Fig.~\ref{fig:rec_mapping}), was first introduced by Shannon in his seminal 1949 paper~\cite{Shannon49}. Later work has extended this mapping to a \emph{spiral type} as well as to N:1 mapping~\cite{Brante13}. AJSCC achieves optimal performance in rate-distortion ratio, whereas to achieve such optimality using \emph{separate} source and channel coding, complex encoding/decoding and long block-length codes would be required, causing delays and energy inefficiencies. Shannon mapping has the two-fold property of (1) compressing the sources (by means of N:1 mapping) and (2) being robust to wireless channel distortions as the noise only introduces errors along the parallel lines (or the spiral curve). In contrast, linear mapping techniques such as Quadrature Amplitude Modulation (QAM) have errors spread on the entire constellation plane. Therefore, channel noise has less effect on the error performance for Shannon mapping as compared with linear modulation schemes. AJSCC requires simple compression and coding, and low-complexity decoding. In rectangular-type mapping, to compress the source signals (``sensing source point"), say Humidity and Temperature voltages ($V_H$, $V_T$), the point on the space-filling curve with minimum Euclidean distance from the source point is chosen (``AJSCC mapped point"), as illustrated in Fig.~\ref{fig:rec_mapping}, via a simple projection on the curve. The transmitted signal is then the ``accumulated length'' of the lines from the origin to the mapped point, where the error introduced by the mapping is controlled by the spacing $\Delta_H$ between lines.

\textbf{Existing AJSCC Circuits: }Existing AJSCC solutions use \emph{all-digital hardware}, and are power demanding and of high circuit complexity. For example, a Software-Defined Radio~(SDR) system to realize AJSCC mapping has been reported in~\cite{Garcia11}. The mapping was also recently implemented in an optical digital communication system in~\cite{Romero14}. Shannon-mapping encoding was adopted in~\cite{stopler14} for a digital video transmission. No existing work has implemented AJSCC using an all-analog, low-complexity design, which is indeed needed for critical futuristic applications requiring low-power sensing such as persistent wireless sensor networks and Internet of Things~(IoTs), where the sensing circuit is expected to occupy small space, consume very low power, and be inexpensive.

\textbf{Our Proposal: }We are proposing a novel all-analog circuit to realize AJSCC and to achieve the objective of very low circuit complexity and power consumption. The proposal is based on the observation that the voltage of each level can be generated by Voltage Controlled Voltage Sources~(VCVS), which output a voltage that is proportional to the input voltage; also, these sources can be switched on or off based on the input signal. Our contributions can be summarized as follows:
\begin{itemize}
  \item Proposing the first \emph{all-analog circuitry} for AJSCC for low-complexity and low-power applications;
  \item Verifying the circuitry and assessing its performance via Spice simulations as well as by prototype development via breadboard and Printed Circuit Board~(PCB) hardware implementations.
\end{itemize}

\begin{figure}
\begin{center}
\includegraphics[width=3in]{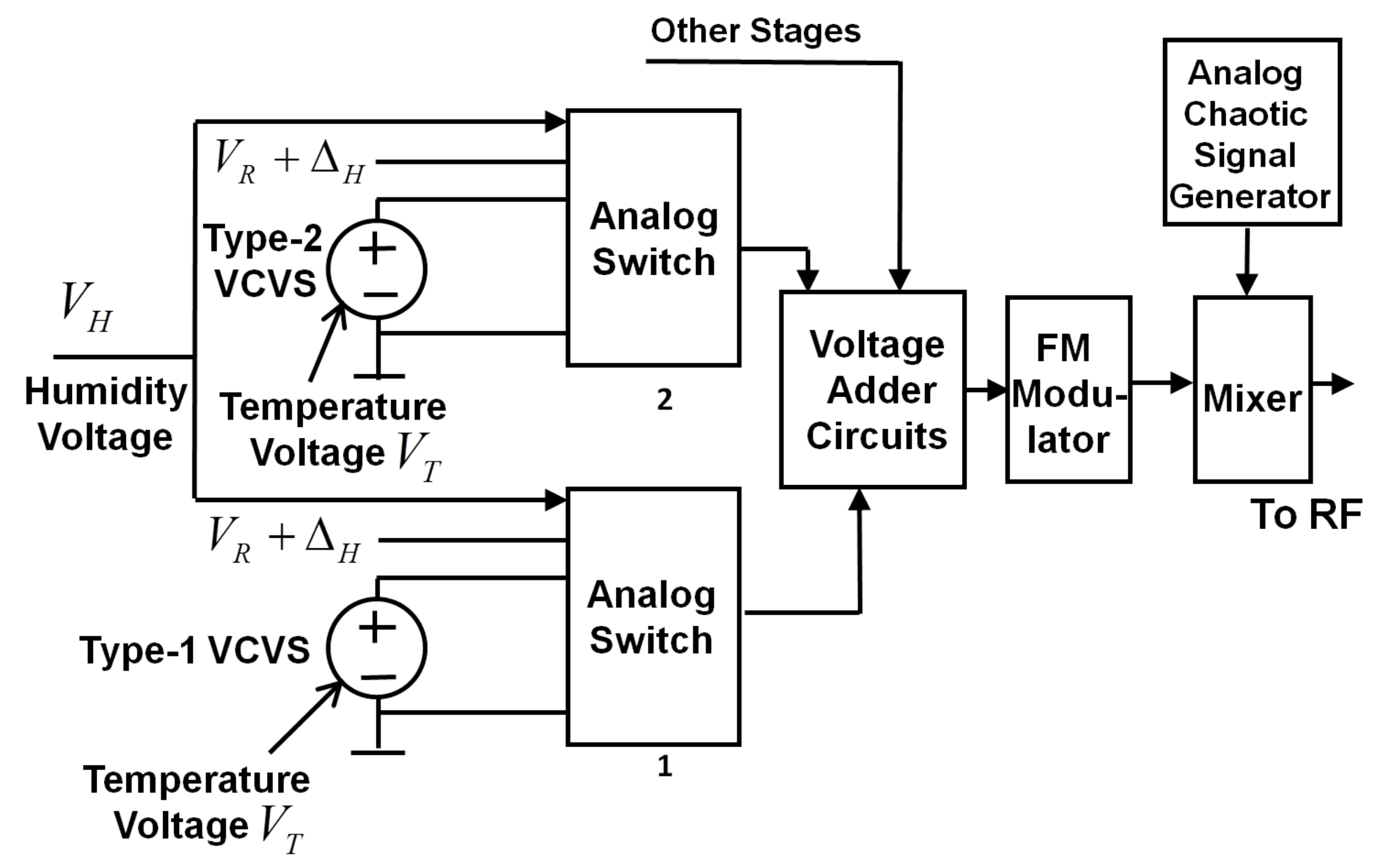}
\end{center}
\caption{\textbf{Proposed Analog Circuit for Shannon's Rectangular Mapping (only the first stage is shown)}. $V_H$ in comparison with threshold voltages generates select signals for the two analog multiplexers to decide which of the three inputs goes to the output. The outputs of both muxes are added to give this stage output. Similar outputs from higher stages are added to give the AJSCC encoded voltage, which is then FM modulated and mixed with semi-orthogonal codes before the Radio Frequency (RF) transmission.}\label{fig:prop_ckt}
\end{figure}

\textbf{Paper Outline: }In Sect.~\ref{sec:prop_ckt}, we present our all-analog circuit realization of AJSCC; then, in Sect.~\ref{sec:circuit}, we provide the circuit performance results from both Spice simulations and implementations; finally, in Sect.~\ref{sec:conc}, we draw the conclusions.

\section{All-analog Circuit Realization of AJSCC}\label{sec:prop_ckt}

Our proposed analog circuit is depicted in Fig.~\ref{fig:prop_ckt}. The VCVS accepts voltage as input, and outputs a voltage that is a function of the input voltage. By observing the rectangular-mapping curve (see Fig.~\ref{fig:rec_mapping}), there are two types of output increments. In the first (which we call Type 1), the output $V_O$ increases linearly with increasing $V_T$, i.e., $V_{O,Type1} \propto V_T$. This happens when we traverse the curve from left to right on odd-numbered lines to reach the mapped point. In the second (Type 2), the output decreases linearly with increasing $V_T$, i.e., $V_{O,Type2}=V_R-V_{O,Type1}$, which corresponds to traversing the curve from right to left on even-numbered lines to reach the mapped point. We realize these two types of outputs using two types of VCVS, where the overall mapping output is the voltage summation of the activated VCVS blocks, which represents the accumulated length of the curve from the origin to the mapped point. Note that $V_H$ controls which VCVS levels/lines are activated, while $V_T$ controls the VCVS output, as mentioned above. A stack of analog switches is used to implement the former and also to control the output of all levels. The number of switches is determined by the mapping resolution sought ($\Delta_H$).

The first and second levels of the circuit are shown in Fig.~\ref{fig:prop_ckt}, where the output of each level is controlled by the corresponding analog switch. Each switch has three inputs: $V_R+\Delta_H$, $GND$, and the output of either a Type-1 or Type-2 VCVS. $V_R$ is a fixed voltage proportional to the x-axis range of the curve. Odd or even numbered switches are connected to Type-1 or Type-2 outputs, respectively. The number of activated switches is determined by considering both $V_H$ and the chosen resolution $\Delta_H$. This also determines the activation voltage of each switch.

The switching logic is explained as follows: each switch is activated only if $V_H$ is greater than a certain value based on the switch's level in the stack. If not activated, the output of the switch is connected to $GND$; if activated, the output is equal to the VCVS output within a certain range ($\Delta_H$) of $V_H$ above the activation voltage. Once $V_H$ exceeds this range, the switch outputs its saturation voltage of $V_R+\Delta_H$. Finally, a voltage adder adds the outputs of all the (activated) switches and sends this analog signal to the FM analog circuit for Radio Frequency (RF) transmission. We refer the two levels, consisting of a Type-1 and Type-2 circuitry, as a \emph{stage}. Figure~\ref{fig:prop_ckt} represents the first stage, which performs the mapping for the first two parallel lines (from the bottom). 

In summary, if the mapped point lies on an odd-numbered line i.e., if $V_H$ is mapped to an odd-numbered level $n$, then the output of the voltage adder will be $V_O=(n-1)(V_R+\Delta_H)+V_T$; conversely, for an even level, the output will be $V_O=(n-1)(V_R + \Delta_H)+(V_R-V_T)$. As an example, for the highlighted point in Fig.~\ref{fig:rec_mapping}, the output of the adder will be $3(V_R+\Delta_H)+(V_R-V_T)$. A Sallen-Key structure or a simple voltage divider can be used for Type-1 VCVS along with a subtractor for Type-2 VCVS, comparators can be used to detect $V_H$ range, and a simple multiplexer for analog switching. %

\section{Circuit Performance Evaluation}\label{sec:circuit}
To verify our design and to assess the performance of our AJSCC mapping circuit, we have carried out thorough Spice simulations as well as performed prototype development via breadboard and PCB hardware implementations.

\textbf{Spice Simulations: }The AJSCC mapping circuit has been simulated in Spice. As mentioned in~Sect.~\ref{sec:prop_ckt}, our circuit consists of multiple identical stages stacked one on top of the other (i.e., hardware is duplicated). Each such stage consists of two levels -- Type-1 and Type-2 VCVS-based circuitry. We considered a temperature sensor ($AD22100$) and humidity sensor ($HIH4000$) having range, respectively, $1.375-3.625~\mathrm{V}$ for $0-100 ^{\circ}C$, and $0.8-3.8~\mathrm{V}$ for $0-100 \%$. We first subtract the non-zero offset from $V_T$'s range as it adds to the total mapped length without carrying information. We have also removed the offset from $V_H$ as it simplifies the design; consequently, the offset-removed voltages, $V_{T0}$ and $V_{H0}$, range in $[0,2.25]$ and $[0,3.0]~\mathrm{V}$, respectively. Just for simulation and implementation purposes, we considered a $10\%$ resolution in humidity giving us $\Delta_H=0.3~\mathrm{V}$. Since it is a second stage, Type 1 is in linear region (and Type 2 is OFF) when $0.45~\mathrm{V} < V_{H0} \le 0.75~\mathrm{V}$ and Type 2 is in linear region (and Type 1 is in saturation) when $0.75~\mathrm{V} < V_{H0} \le 1.05~\mathrm{V}$. These voltages are also shown in Fig.~\ref{fig:rec_mapping} for ease of understanding.

\begin{figure}
\begin{center}
\includegraphics[width=4.35in]{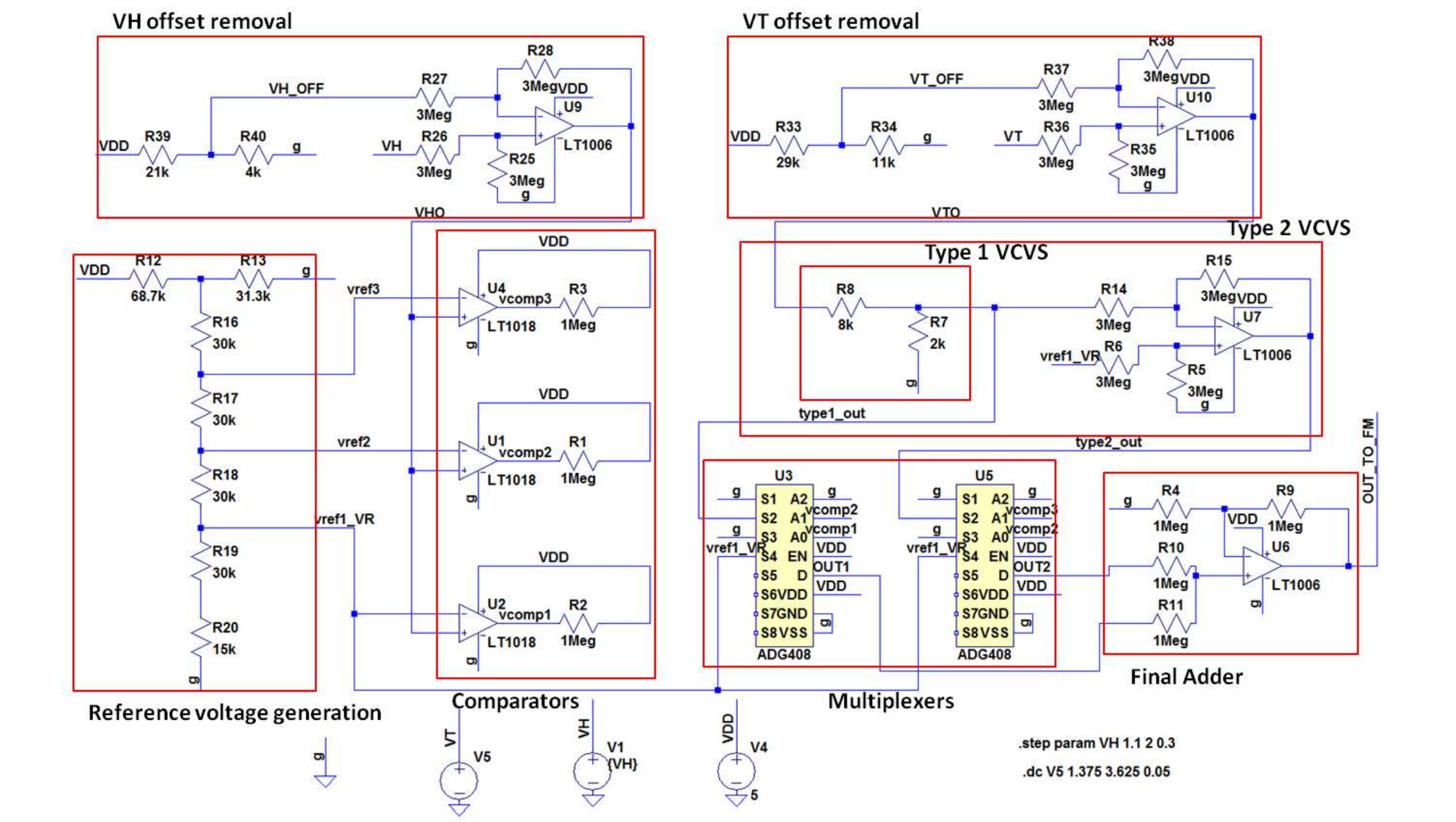}
\end{center}
\caption{\textbf{LTSpice schematic} of analog components of the second stage of our AJSCC-mapping analog circuit, where the functional blocks are identified.}\label{fig:spice_ckt}
\end{figure}

\begin{figure}
\begin{center}
\includegraphics[width=2.5in]{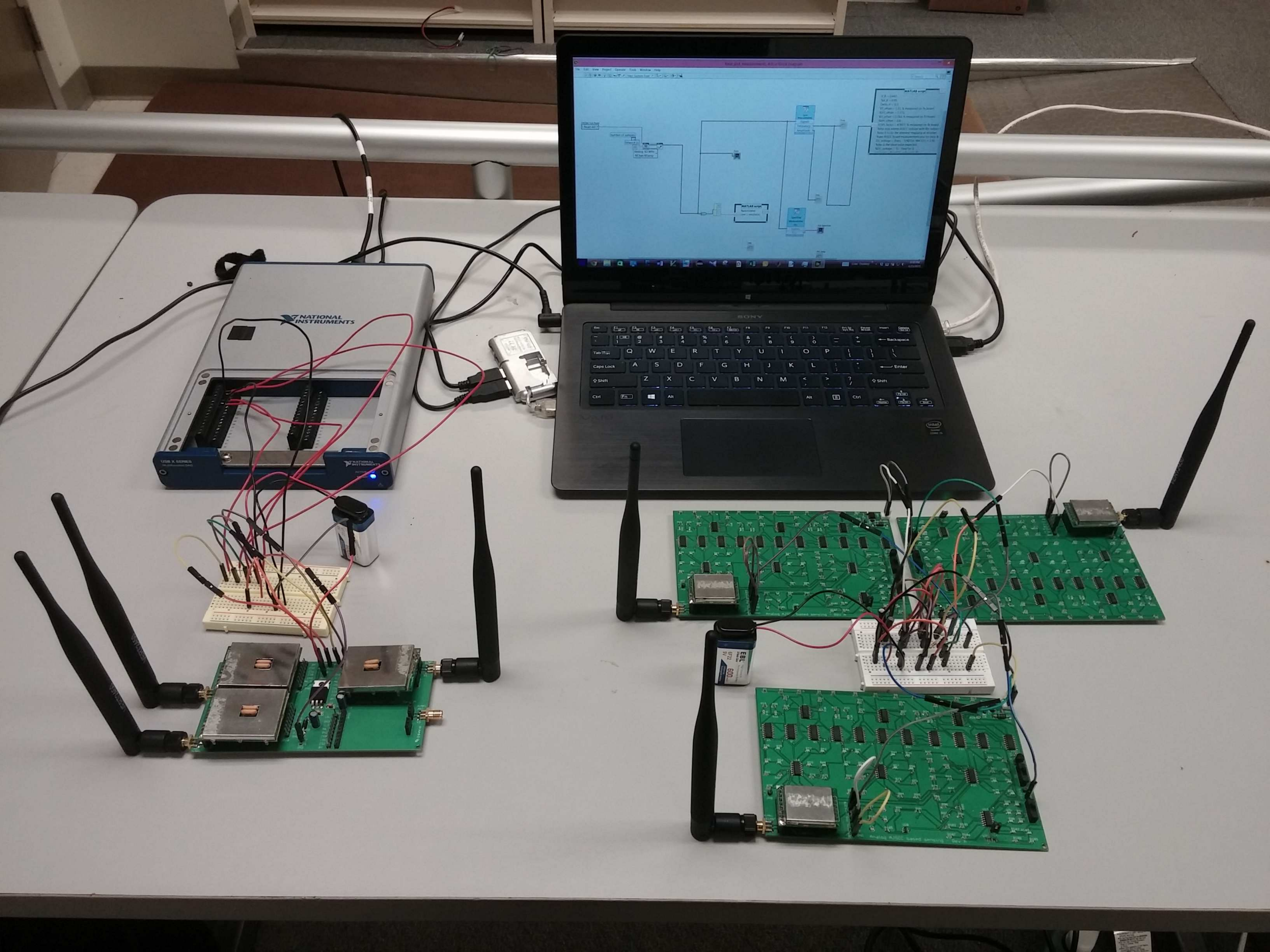}
\end{center}
\caption{\textbf{System prototype. }Three analog sensors (right bottom) communicating to a receiver (left bottom). The baseband signals of all three channels are captured using NI Digital Acquisition System (DAQ) and processed/decoded on host computer using LabView/MATLAB.}\label{fig:prototype}
\end{figure}

Figure~\ref{fig:spice_ckt} shows the detailed schematic of the second stage of our circuit simulated in LTSpice, a Spice simulation tool. While we have prototyped the entire system (Fig.~\ref{fig:prototype}), we have chosen to simulate only the second stage as the $V_{H0}$ mapping range of both levels is more general compared to the first stage where the Type 1 has half of this value (i.e., $0$ to $0.15~\mathrm{V}$). Offset removal is done using an OpAmp-based subtractor circuit. In fact, we have used well-known OpAmp-based subtractors and non-inverting adders for all subtractor and adder functionalities.  Both $V_T$ and $V_H$ offset-removal blocks are indicated in the schematic. $V_{H0}$ range comparison is achieved using comparator ICs. Reference voltages ($0.45$, $0.75$, and $1.05~\mathrm{V}$) for comparison are generated using a simple voltage-divider circuit, as shown in the figure. In each level (Type 1 or Type 2), the outputs of the comparators act as select lines for the analog multiplexer (switches), which receives three inputs, i.e., saturation voltage $V_R$, Type-1 VCVS output ($type1\_out$) for level 1 or Type-2 VCVS output ($type2\_out$) for level 2, and ground signal ($g$). Saturation voltage, $V_R$, is the output of Type-1 VCVS when the sensor senses the maximum temperature, i.e., when $V_{T0}=2.25~\mathrm{V}$. $V_R$ and $g$ signals are passed to the output of a Type-1 multiplexer when $V_{H0} \le 0.45~\mathrm{V}$ and $V_{H0} > 0.75~\mathrm{V}$, respectively. Similar statements hold for a Type-2 multiplexer. We have mentioned above the condition for $V_{H0}$ when Type-1 (or Type-2) outputs are passed to the multiplexer. Type-1 VCVS is implemented as a simple voltage-divider circuit with 1:5 ratio, which means $V_R=0.45~\mathrm{V}$ as $V_{T0,max}=2.25~\mathrm{V}$. The output of Type 1 is subtracted from $V_R$ to get the Type-2 output.

Lastly, the outputs of both multiplexers are added to give the final output of this second stage. Similar outputs from the other stages are added to be given as input to the FM modulator.
$V_R=V_1$ is determined by the maximum allowed voltage input to the FM modulator.
We used manufacturer-provided Spice models to capture real performance; on the other hand, we were also limited by them as there are very few such models. Linear Technology's $LT1006$ and $LT1018$, and Analog Devices' $ADG408$ have been used, respectively, for OpAmp, comparator, and analog multiplexer functionalities.

\begin{figure*}[ht!]
\centering
\begin{tabular}{ccc}
\hspace{-0.6cm}
\includegraphics[width=0.34\textwidth]{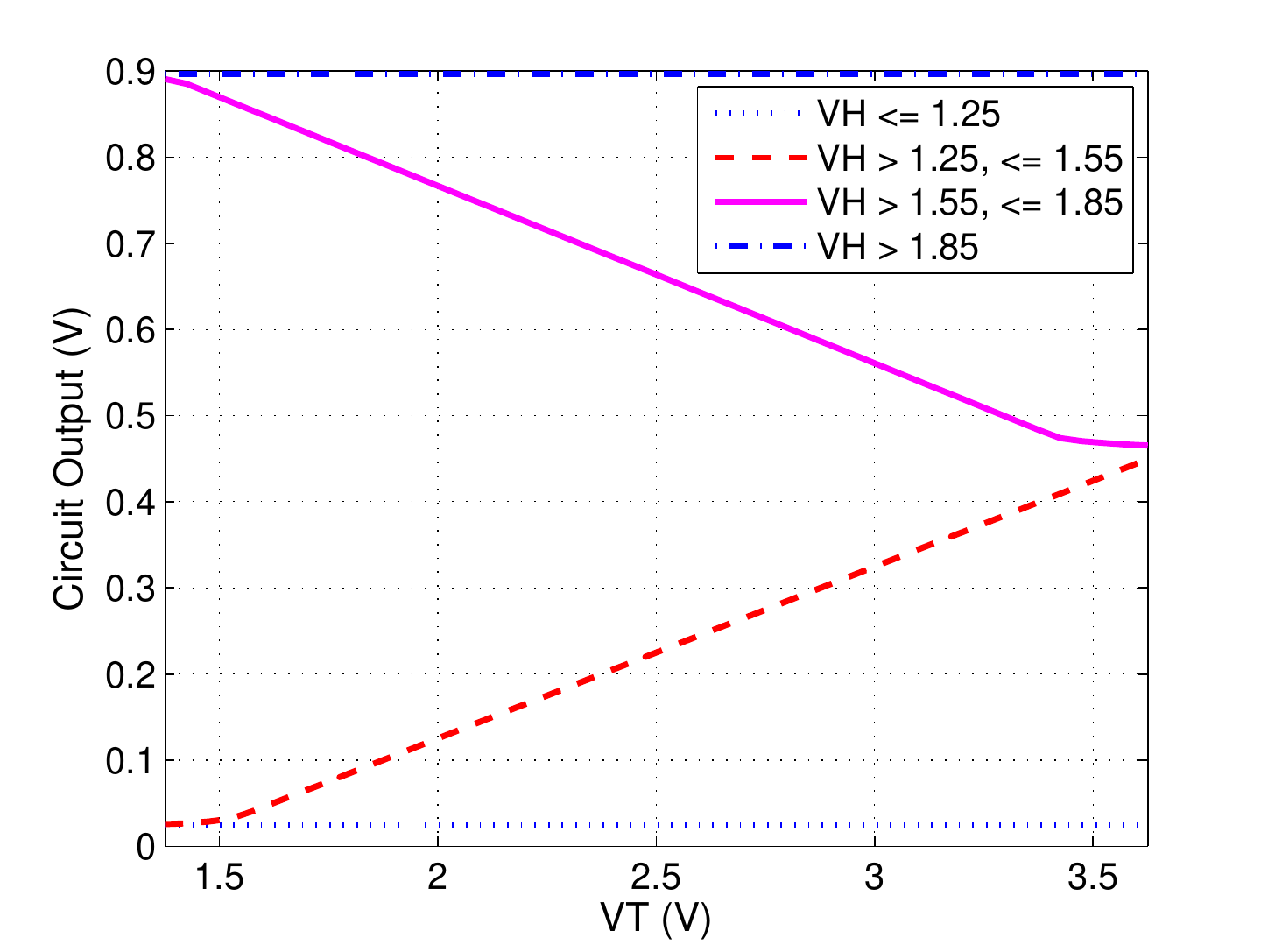} & \hspace{-0.5cm}
\includegraphics[width=0.34\textwidth]{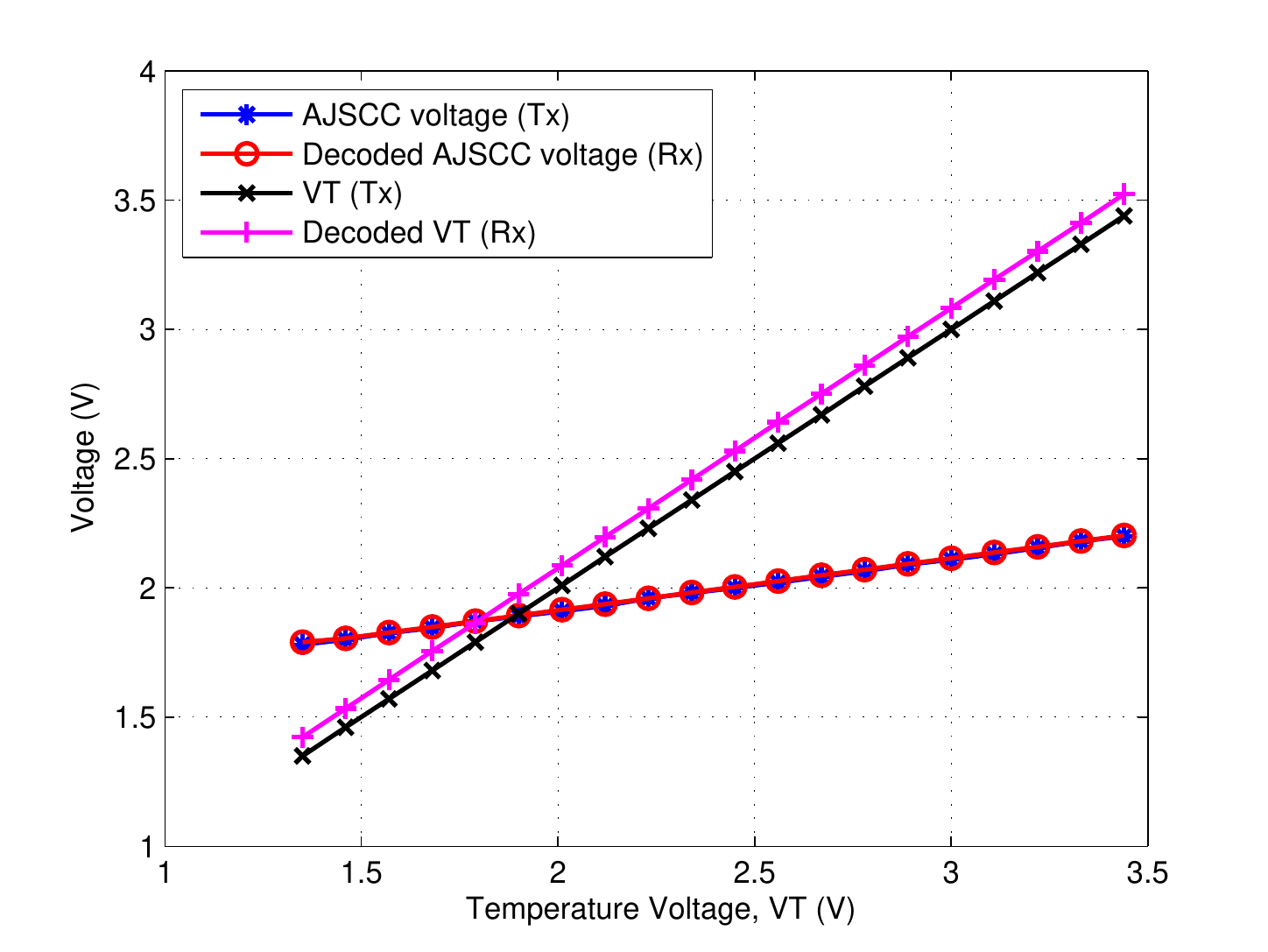} & \hspace{-0.5cm}
\includegraphics[width=0.34\textwidth]{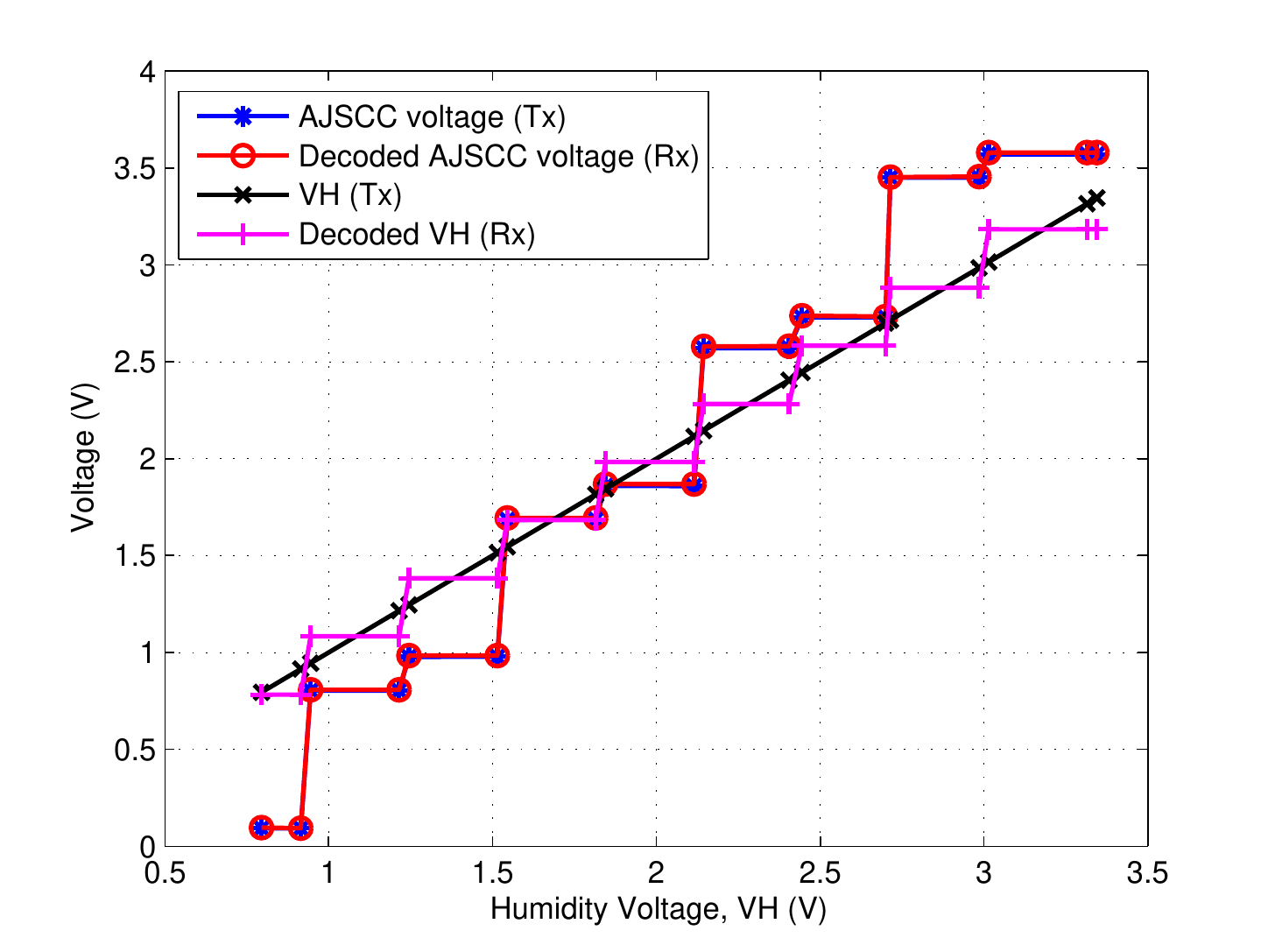}  \\
\hspace{-0.6cm}
\small (a)   & \hspace{-0.5cm} \small(b)    & \hspace{-0.5cm} \small(c)
\end{tabular}
\caption{(a) Breadboard AJSCC output of the \emph{second stage}; notice the four different regions -- off, linear (Type-1 VCVS), linear (Type-2 VCVS), and saturation. (b) PCB AJSCC output for varying $V_T$ with $V_H=1.83~\mathrm{V}$, note that the AJSCC output is linear with increasing $V_T$ and that $\hat V_T$ closely follows $V_T$ (average $\hat V_H$ is $1.98~\mathrm{V}$). (c) PCB AJSCC output for varying $V_H$ with $V_T=1.71~\mathrm{V}$; note that the AJSCC output and $\hat V_H$ are discrete (average $\hat V_T$ is $1.72~\mathrm{V}$).}\label{fig:hw_results}
\end{figure*}

Figure~\ref{fig:hw_results}(a) shows the output of second stage over its entire mappable range of $V_T$ and $V_H$. It can be seen that the output is almost zero when $V_H$ is below $1.25~\mathrm{V}$ and that, when $1.25 < V_H \le 1.55$, Type 1 mainly drives the output (with zero contribution from Type 2); conversely, when $1.55 < V_H \le 1.85$, Type 2 mainly drives the output (with $V_R$ contribution from Type 1); finally, when $V_H > 1.85$, the output is $2V_R$ (i.e., $V_R$ from each VCVS). Note that these are the actual sensor output voltages and not offset-removed values. By this manner, this circuit captures and maps the sensor outputs in the range $1.25 \le V_H < 1.85$ and $1.375 \le V_T \le 3.625$. We can observe some saturation behaviors (output not being exactly zero) when both levels are inactive and at the beginning of Type-1 and Type-2 active modes, which we attribute to the saturation of the OpAmps.

\textbf{Breadboard and PCB Hardware Implementations: }We have also implemented this single stage on a breadboard to verify the reproducibility of the simulation results using real hardware. We found that the results were vey similar to our simulation counterparts, as depicted in Fig.~\ref{fig:hw_results}(a), with less saturation effects in fact. Breadboard results motivated us to go even a step further and implement the full circuit (all stages) along with the RF part (as a COTS component) i.e., a full-fledged sensor on a PCB entirely designed by us (Fig.~\ref{fig:prototype}). This sensor consists of three major blocks -- AJSCC encoding, DC-to-sine wave conversion circuit, and a RFIC module. The AJSCC encoding block takes temperature and humidity voltages as input, and outputs the AJSCC-encoded voltage. It implements 11 VCVS levels in total, as per the setup described above ($\Delta_H=0.3~\mathrm{V}$). The DC-to-sine wave block converts the AJSCC-encoded DC voltage to a sine wave, which will then be FM modulated and RF transmitted by the RFIC module. We have also designed a receiver consisting of a RFIC module and a NI Digital Acquisition~(DAQ) system, and used LabView to process the decoded $V_T$ and $V_H$ values.

Figure~\ref{fig:hw_results}(b) shows the sensor's AJSCC-mapped voltage for varying $V_T$ (with $V_H$ held constant at $1.83~\mathrm{V}$), while Fig.~\ref{fig:hw_results}(c) shows it for varying $V_H$ (with $V_T$ held constant at $1.71~\mathrm{V}$). As expected, when $V_H$ is fixed, we observe a linear increase in the AJSCC output as $V_T$ is increased within its range. This is because of the linear VCVS functionality. However, a step output is expected when $V_T$ is fixed while $V_H$ is varied within its range. This is because all those $V_H$ values between two threshold-level voltages are mapped to the same voltage. It can also been seen in both cases that the decoded AJSCC voltage is very close to the transmitted AJSCC voltage, i.e., $\hat V_T$ closely follows $V_T$, while, as expected, a step is observed in $\hat V_H$.
\begin{figure}
  \begin{center}
 \includegraphics[width=2.6in]{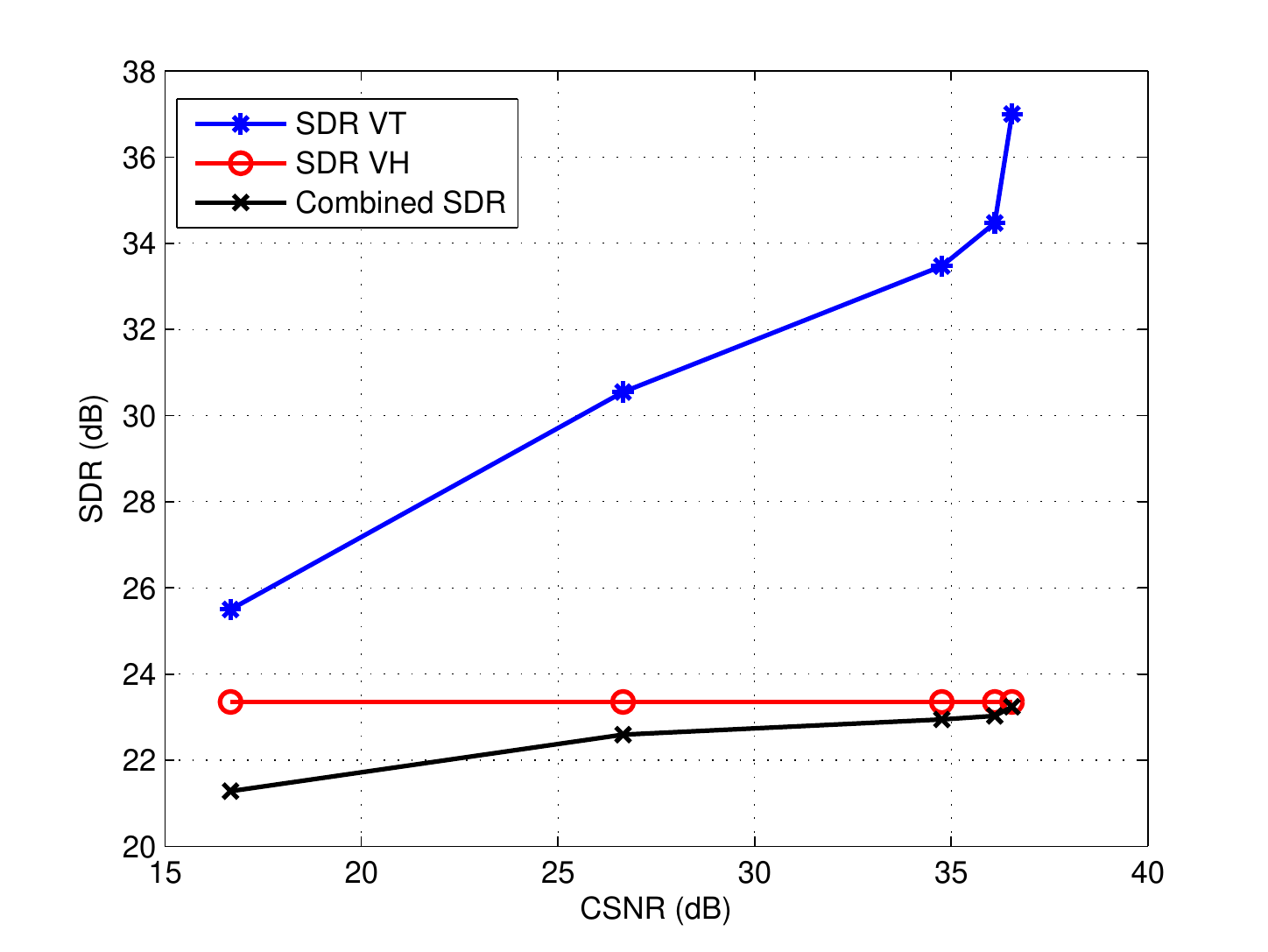}
 \end{center}
 \caption{SDR-vs-CSNR performance when three analog sensors communicate simultaneously using AJSCC to a digital Cluster Head (CH), the receiver.}
 \label{fig:csnr_sdr_tx}
\end{figure}

We measured our prototype's performance when three sensors (Tx) communicate simultaneously to a digital Cluster Head (Rx) using Frequency Division Multiple Access~(FDMA). Figure~\ref{fig:csnr_sdr_tx} plots the SDR vs CSNR, where the former is the Signal-to-Distortion Ratio, i.e., the inverse in logarithmic scale of the Mean Square Error (MSE); and the latter is the Channel Signal-to-Noise Ratio, i.e., the SNR of the baseband signal at the output of the receiver RF module. SDR and CSNR are plotted by varying the Tx-Rx distance (with $V_T$ and $V_H$ fixed). As CSNR increases the SDR also increases, indicating the improved signal reconstruction performance.

\textbf{Power and Cost Considerations: }State-of-the-art sensor nodes consume $\approx0.5~\mathrm{mA}$ in active mode and a few $\mathrm{\mu A}$ in sleep mode with supply voltages in the range $1.8-3.0~\mathrm{V}$, i.e., $0.9-1.5~\mathrm{mW}$ without taking into account the radio power: the active power consumption is mainly due to the microcontroller and Analog-to-Digital (A/D) conversions. In contrast, our all-analog sensor does not use power-hungry A/D's or microcontrollers. The current drawn by the AJSCC baseband circuitry (using COTS OpAmps, Multiplexers, etc.), i.e., the entire board excluding the RFIC module, is $\approx3~\mathrm{mA}$ with a supply voltage of $5~\mathrm{V}$ (equivalent to $\approx15~\mathrm{mW}$); the cost of the AJSCC PCB is about $\$25$. These numbers, which are high because of (1) the use of COTS components and (2) duplication of hardware for each stage, can be reduced \emph{drastically} if Integrated Circuit~(IC) design is adopted. While our implementation serves as feasibility study, we believe the power consumption can be reduced to $\approx150~\mathrm{\mu W}$ if our circuit is redesigned using the latest nm-technology components (for OpAmps, Comparators, and Multiplexers). 

Let us provide a rough estimate: our circuit in total (5 and half stages/11 levels) consists of 16 OpAmps, 17 Comparators, and 11 Multiplexers, where OpAmps are clearly the major contributors to the overall power consumption. There are many low-power designs proposed for these components. For example, a low-power OpAmp~\cite{opamp} consuming about $8~\mathrm{\mu W}$, a comparator~\cite{comparator} consuming about $12.7~\mathrm{nW}$, and an analog multiplexer ($ADG704$) consuming about $10~\mathrm{nW}$ can be used for our circuit resulting in a power consumption of $\approx\mathrm{130 \mu W}$. We are also optimistic that the sensor cost would reduce to less than $\$5$ leveraging economies of scale via mass production using the latest IC technology. Achieving both goals will enable critical futuristic applications such as persistent wireless sensing and IoT-based solutions.

\section{Conclusion and Future Work}\label{sec:conc}
To achieve low-power, low-complexity compression of sensor signals into one with controlled distortion, while also being robust against wireless channel impairments, we proposed a novel all-analog circuit for rectangular-type Analog Joint Source Channel Coding~(AJSCC) using Voltage Controlled Voltage Sources~(VCVS). The proposal is evaluated by Spice simulations as well as breadboard and PCB implementations. We will address the hardware duplication issue as future work.

\textbf{Acknowledgments: }We thank Rutgers ECE graduate student Priyank Bharad for helping with the implementation.

\bibliographystyle{IEEEtran}\small
\bibliography{MyPublications_v1,career_v1,sensor_v4,references_v3.0,cross-layer_v3,underwater_v19,RobustConsensus}

\begin{thebibliography}{1}
\providecommand{\url}[1]{#1}
\csname url@rmstyle\endcsname
\providecommand{\newblock}{\relax}
\providecommand{\bibinfo}[2]{#2}
\providecommand\BIBentrySTDinterwordspacing{\spaceskip=0pt\relax}
\providecommand\BIBentryALTinterwordstretchfactor{4}
\providecommand\BIBentryALTinterwordspacing{\spaceskip=\fontdimen2\font plus
\BIBentryALTinterwordstretchfactor\fontdimen3\font minus
  \fontdimen4\font\relax}
\providecommand\BIBforeignlanguage[2]{{%
\expandafter\ifx\csname l@#1\endcsname\relax
\typeout{** WARNING: IEEEtran.bst: No hyphenation pattern has been}%
\typeout{** loaded for the language `#1'. Using the pattern for}%
\typeout{** the default language instead.}%
\else
\language=\csname l@#1\endcsname
\fi
#2}}

\bibitem{Hekland05}
F.~Hekland, G.~Oien, and T.~Ramstad, ``Using 2:1 {S}hannon mapping for joint
  source-channel coding,'' in \emph{Data Compression Conference}, March 2005,
  pp. 223--232.

\bibitem{Fresnedo15}
O.~Fresnedo, J.~Gonzalez-Coma, M.~Hassanin, L.~Castedo, and J.~Garcia-Frias,
  ``Evaluation of analog joint source-channel coding systems for multiple
  access channels,'' \emph{IEEE Transactions on Communications}, vol.~63,
  no.~6, pp. 2312--2324, June 2015.

\bibitem{Shannon49}
C.~Shannon, ``Communication in the presence of noise,'' \emph{Proceedings of
  the IRE}, 1949.

\bibitem{Brante13}
G.~Brante, R.~Souza, and J.~Garcia-Frias, ``Spatial diversity using analog
  joint source channel coding in wireless channels,'' \emph{IEEE Transactions
  on Communications}, vol.~61, no.~1, pp. 301--311, January 2013.

\bibitem{Garcia11}
J.~Garcia-Naya, O.~Fresnedo, F.~Vazquez-Araujo, M.~Gonzalez-Lopez, L.~Castedo,
  and J.~Garcia-Frias, ``Experimental evaluation of analog joint source-channel
  coding in indoor environments,'' in \emph{IEEE International Conference on
  Communications (ICC)}, June 2011, pp. 1--5.

\bibitem{Romero14}
S.~Romero, M.~Hassanin, J.~Garcia-Frias, and G.~Arce, ``Analog joint source
  channel coding for wireless optical communications and image transmission,''
  \emph{Journal of Lightwave Technology}, vol.~32, no.~9, pp. 1654--1662, May
  2014.

\bibitem{stopler14}
D.~Stopler, ``Device method and system for communicating data,'' Jan 2014, {US}
  Patent 8,625,709.

\bibitem{opamp}
W.~S. Y.~Libin, M.~Steyaert, ``A 0.8v 8µw cmos ota with 50-db gain and 1.2-mhz
  gbw in 18-pf load,'' June 2003, pp. 297--300.

\bibitem{comparator}
A.~Valaee and M.~Maymandi-Nejad, ``An ultra low-power low-voltage track and
  latch comparator,'' in \emph{Electronics, Circuits, and Systems (ICECS), 2010
  17th IEEE International Conference on}, Dec 2010, pp. 186--189.

\end{thebibliography}

\end{document}